# Two-phase argon and xenon avalanche detectors based on Gas Electron Multipliers

A. Bondar, A. Buzulutskov [*], A. Grebenuk,
D. Pavlyuchenko, R. Snopkov, Y. Tikhonov

*Budker Institute of Nuclear Physics, 630090 Novosibirsk, Russia*

**Abstract**

We study the performance of two-phase avalanche detectors based on Gas Electron Multipliers (GEMs) and operated in an electron-avalanching mode in Ar and Xe. Emission, gain, energy resolution and stability characteristics of the detectors were studied. Rather high gains, reaching 5000, and stable operation for several hours were observed in the two-phase Ar avalanche detector using a triple-GEM multiplier. The signals induced by X-rays, β-particles and γ-rays were successfully recorded. Preliminary results were obtained in the two-phase Xe avalanche detector: the maximum gain of the triple-GEM in two-phase Xe and Xe+2%$CH_4$ was about 200. The results obtained are relevant in the field of two-phase detectors for dark matter searches, coherent neutrino scattering, PET and digital radiography.

## 1. Introduction

The studies of two-phase (liquid-gas) detectors operated in an electron-avalanching mode (Two-Phase Avalanche Detectors) have been motivated by a growing interest in their applications in cryogenic experiments where the primary ionization signal is weak: in coherent neutrino scattering [1], dark matter searches [2], solar neutrino detection [3] and Positron Emission Tomography (PET) [4]. A typical deposited energy in such experiments is in the range of 1-500 keV, i.e. might be rather low. Accordingly, the signal should be amplified.

In two-phase avalanche detectors [5] (see Fig. 1) the primary electrons produced in the noble liquid are emitted into the gas phase by an electric field where they are multiplied in saturated vapor above the liquid using a gas multiplier, in particular the Gas Electron Multiplier (GEM) [6]. In some cases, the primary scintillation signal should also be detected, to reduce the background by comparing the ionization and scintillation signals, in dark matter searches [2], and to provide fast coincidences between collinear gamma quanta, in PET [4]. It was suggested [5,7] that the multi-GEM multiplier would provide the detection of both the ionization and scintillation signals in two-phase avalanche detectors (Fig. 1). The latter would be achieved by depositing an UV-sensitive photocathode, namely CsI, on top of the first GEM. This technique has been recently developed for GEM-based gas photomultipliers [8]. To provide high detection efficiency for scintillations and direct ionization, the multi-GEM multiplier should be able to operate at high gains in dense noble gases, in saturated vapor and at cryogenic temperatures. This is a real challenge.

[*] Corresponding author. Email: buzulu@inp.nsk.su

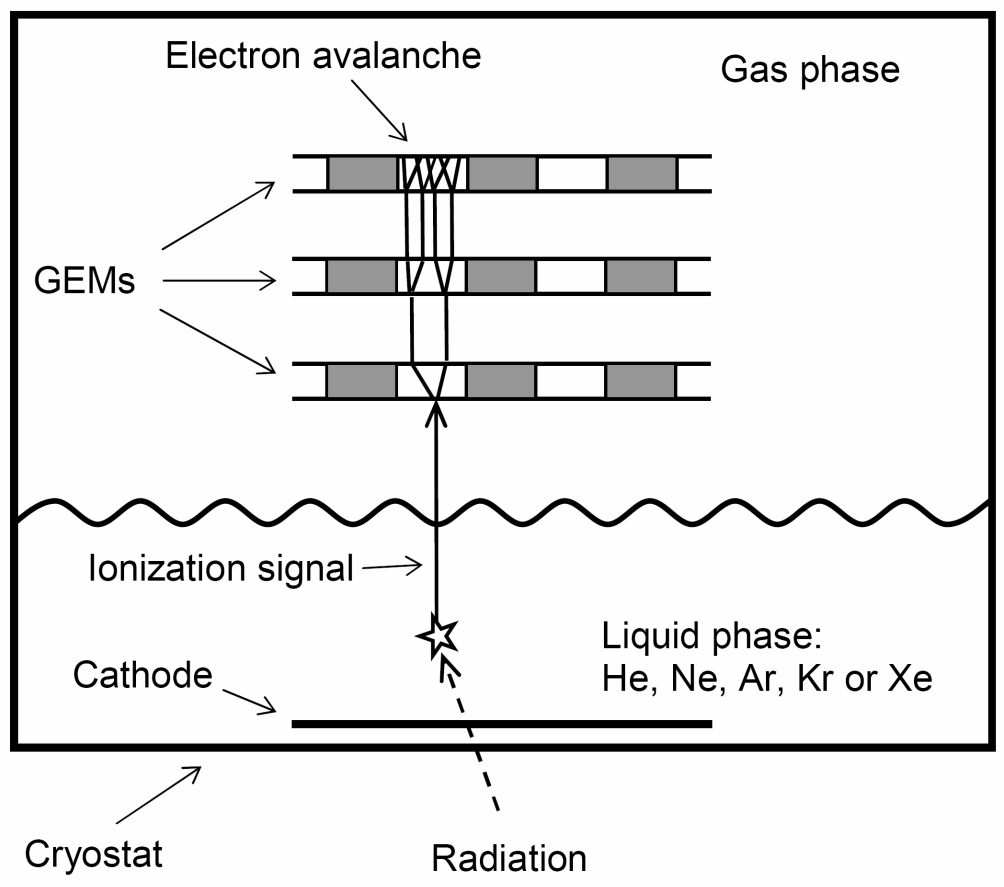

Fig.1 Schematic view of a two-phase avalanche detector based on GEM multipliers.

The remarkable property of GEM structures is that they can operate in all noble gases at reasonable gains [9,10], including at cryogenic temperatures [5,11-13] and even down to 2 K [13]. Later it was shown that other hole-type structures, such as Micro-Hole and Strip Plates [14] and Capillary Plates [15], can also operate in noble gases.

It might be desirable to use all noble gases for different reasons in two-phase detectors. Indeed, the noble liquids proposed for applications in two-phase detectors and optimized for the detector mass and event rate are Ar and Ne for coherent neutrino scattering [1], Xe [2] and Ar [16] for dark matter searches, He and Ne for solar neutrino detection [3], Xe for PET [17], Ar and Kr for digital radiography [18].

On the other hand, the operation of GEM structures in a two-phase mode was studied for Kr-based detectors only [11,12]: the stable operation of a two-phase Kr avalanche detector, at gains reaching several hundreds, was demonstrated for at least 3 hours. Recently, it has also been reported on the Micromegas operation in two-phase Xe [19], but for a short period: in half an hour the signal disappeared. Furthermore, unlike GEM, the Micromegas cannot operate in pure noble gases: therefore a $CH_4$ additive was used there. Since methane dissolves in liquid Xe, this may in principle result in quenching the scintillations and an increase of the electric field needed for efficient electron emission.

Consequently, the studies of the GEM performance in noble gases other than Kr, in a two-phase mode, are of primary importance. In this paper, the performance of GEM structures in two-phase Ar, Xe and Xe+$CH_4$ was studied for the first time. Emission, gain, energy, pulse shape and stability characteristics of two-phase avalanche detectors are presented under irradiation with X-rays, $\gamma$-rays and $\beta$-particles.

## 2. Experimental setup

The results presented in the following sections are obtained in the experimental setup similar to that used for the two-phase Kr detector [11,12]. Three GEM foils, of an active area of 28×28 $mm^2$ each, and a cathode mesh were mounted in a cryogenic vacuum-insulated chamber of a volume of 2.5 l (Fig.1). The distances between the first GEM and the cathode, and between the GEMs, were 6 and 2 mm, respectively. The chamber was cooled by liquid nitrogen using a heat exchanger mounted on the top flange. In the two-phase mode the thickness of the liquid condensate at the chamber bottom was approximately 3 mm, at a vapor pressure of 1 atm. The condensate characteristics, namely its thickness and state (liquid or solid), were monitored by measuring a capacitance of the cathode gap, between the cathode and the first GEM.

The chamber was filled with appropriate gases, Ar or Xe, or gas mixture, Xe+$CH_4$, taken directly from a 40 l bottle connected to the chamber at all times. The electron drift path in liquid Ar before attachment was larger than 3 mm, which is equivalent to an oxygen impurity level of below $10^{-6}$. The electron drift path in liquid Xe was only 0.6 mm (due to impurities released by outgassing and their high solubility in liquid Xe), which is equivalent to an oxygen impurity level of approximately $3\times10^{-6}$.

The detector was operated in either a current or pulse-counting mode. It was irradiated with either continuous or pulsed X-rays from two different X-ray tubes (operated in the current or pulsed mode respectively) through X-ray windows at the chamber bottom having a diameter of 1 cm. The anode voltages of the tubes were typically 30-50 kV. The width and frequency of X-ray pulses were 400 ns (FWHM) and 250 Hz respectively. The detector could also operate in a pulse-counting mode using β-particles from a $^{90}$Sr source and 60 keV γ-rays from a $^{241}$Am source.

The GEM electrodes were biased through a resistive high-voltage divider placed outside the cryostat and similar to that used in Ref. [13]. The GEM multiplier could operate in a single-, double- or triple-GEM configuration, denoted as 1GEM, 2GEM or 3GEM respectively. The avalanche (anode) signal was recorded from the last electrode of the last active GEM element.

The gain value was measured in a pulse-counting mode using pulsed X-rays and a technique described in Ref. [13]: it is defined as the pulse height of the avalanche signal divided by that of the calibration signal. The calibration signal was recorded at the first electrode of the first GEM. Both signals were read out using a charge-sensitive amplifier with a 10 ns rise time and sensitivity of 0.5 V/pC. In addition, in some cases the gain value was measured in a current mode using a technique described in Ref. [11].

Other details of the experimental setup and procedures are described elsewhere [11,12].

## 3. Two-phase Ar avalanche detector

Fig. 2 shows an evolution of the cathode gap capacitance as a function of the Ar gas/vapor pressure during the cooling and heating procedures. During the cooling procedure in the pressure range of 1.3-1.8 atm, the detector operates in a gaseous mode: the capacitance almost does not change and its fluctuations are negligible. Below 1.3 atm the detector operates in a two-phase mode: a liquid Ar condensate is formed at the chamber bottom and both the gap capacitance and its fluctuations are substantially increased. During the heating procedure the liquid condensate at the bottom is preserved up to a vapor pressure of 1.7 atm.

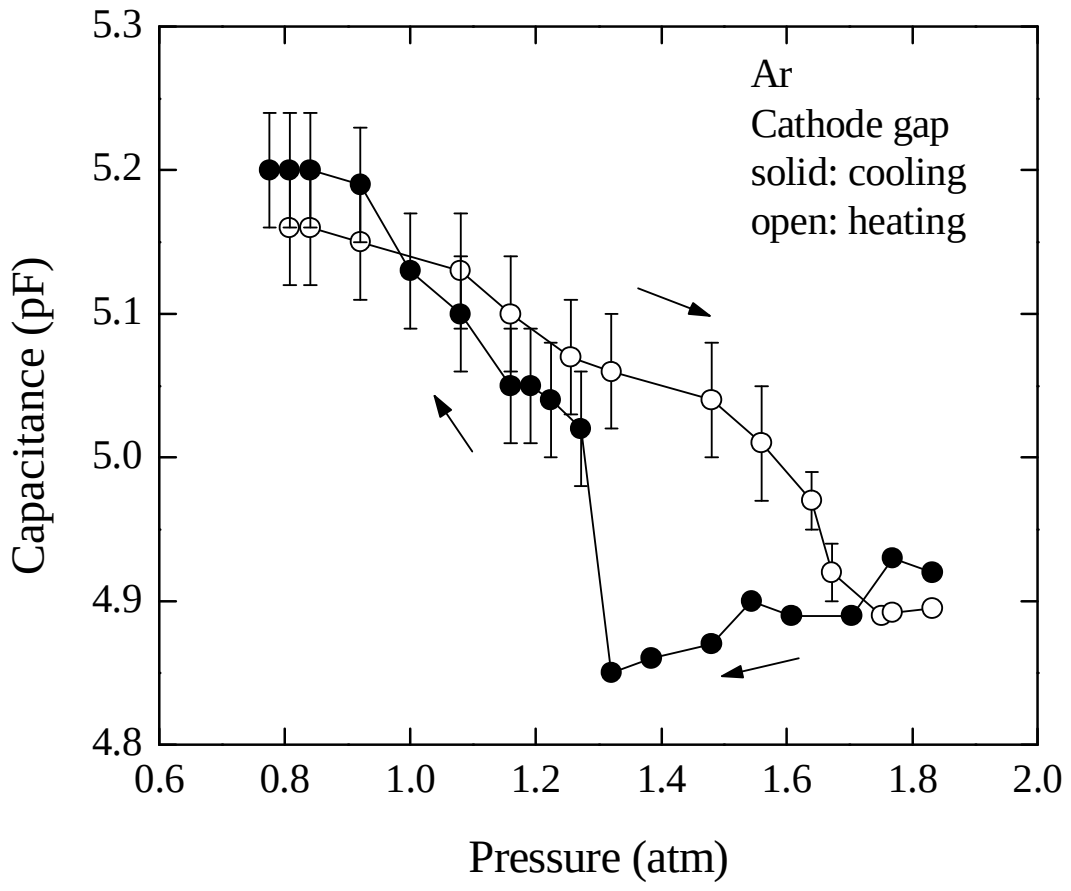

Fig.2 Cathode gap capacitance as a function of the gas/vapor pressure during cooling and heating procedures in Ar.

The capacitance fluctuations are obviously induced by surface waves in the liquid and they are considerably reduced when the condensate is solid (see section 4). It should be remarked that in contrast to the two-phase Kr [12] and Xe detectors, in the Ar detector the condensate at the bottom could not be transformed to a solid state, apparently due to the fact that the boiling point of liquid nitrogen is close to that of liquid argon. Moreover, the last point in the cooling cycle, at 85 K and 0.76 atm, turned out to be the best in terms of the temperature and pressure stabilization: the pressure stability was within 1%. Most of the data presented below were measured at this point.

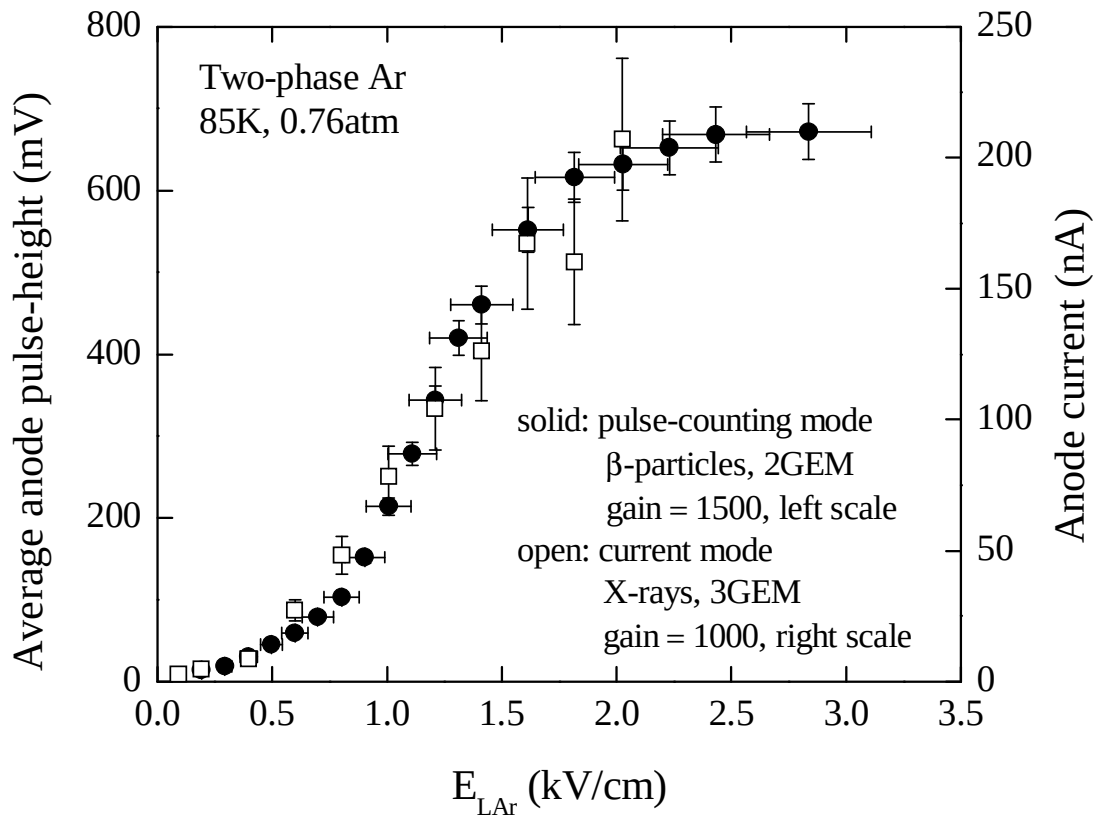

Fig.3 Electron emission characteristics in the two-phase Ar avalanche detector at 85 K and 0.76 atm. Average anode pulse-height from the double-GEM induced by β-particles and anode current from the triple-GEM induced by X-rays are shown as a function of the electric field in liquid Ar. The multi-GEM gains are 1500 and 1000 respectively.

The two-phase Ar detector based on GEM multipliers was successfully operated in an avalanche mode at high gains. Fig. 3 illustrates electron emission through the liquid-gas interface, when the detector is operated either in a pulse-counting or current mode. The anode pulse-height, from the double-GEM, and the anode current, from the triple-GEM, are shown as a function of the electric field in liquid Ar.

One can see that the shapes of the emission curves obtained with the pulse-counting and current techniques are in good correspondence and that the electron extraction efficiency reaches a plateau at a field of 2.5 kV/cm, which is in agreement with the data reported in the literature [20]. It should be remarked that the electric fields needed for efficient electron emission from liquid Ar are by a factor of 2-3 lower than that of liquid Kr and Xe [20], which is a good advantage. The large value of the anode current reached in a current mode should be emphasized: it is 200 nA (right scale in Fig. 3) corresponding to a current density of 2.5 nA/mm$^2$.

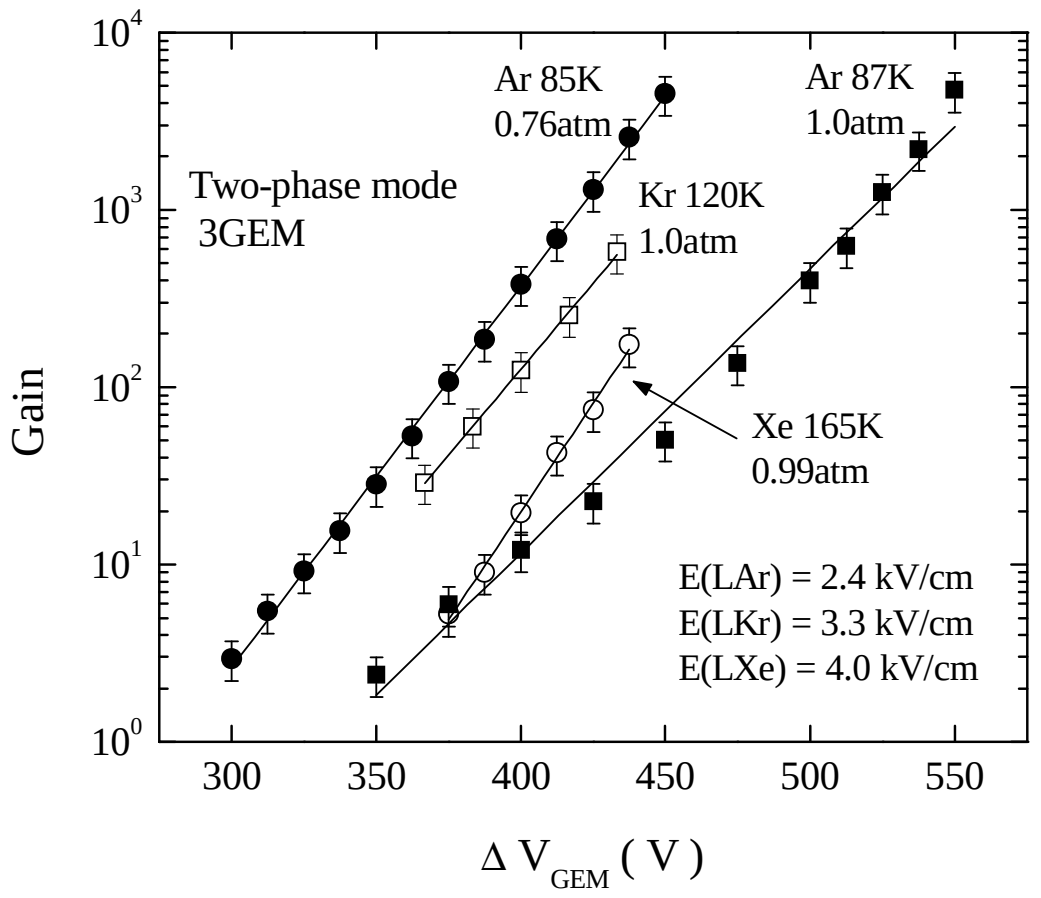

Fig.4 Triple-GEM gain as a function of the voltage across each GEM in two-phase Ar and Xe, measured using pulsed X-rays. The data for two-phase Kr, measured using β-particles, are also shown [11]. The appropriate temperatures, pressures and electric fields in the liquids are indicated. The maximum gains are limited by discharges.

Fig. 4 shows gain-voltage characteristics of the triple-GEM in two-phase Ar at a vapor pressure of 0.76 and 1.0 atm, measured using pulsed X-rays. The maximum gains reached, before the discharge onset, were rather high, about 5000 at both pressures. In the current mode the maximum gain reached was even higher, about $1.2\times10^4$. Gain-voltage characteristics of the single-, double- and triple-GEMs in two-phase Ar are compared in Fig. 5 (here the maximum gains for the single- and double-GEMs were not reached).

One can see that practically for all GEM configurations the gain grows exponentially with voltage. Only at 1.0 atm, there is an indication that the gain starts to grow slightly faster than exponentially, at higher gains. In a current mode, we also did not observe any substantial deviation from an exponential behavior of gain-voltage characteristics (not shown here) even at gains and current densities as high as $10^4$ and 2.5 nA/mm$^2$ respectively. This indicates that charging-up effects are generally not observed in two-phase Ar avalanche detectors. This was not the case for example in two-phase Kr avalanche detectors when operated in a current mode [12]: at gains and anode currents densities exceeding 200 and 100 pA/mm$^2$ respectively the gain started to grow much faster than exponentially.

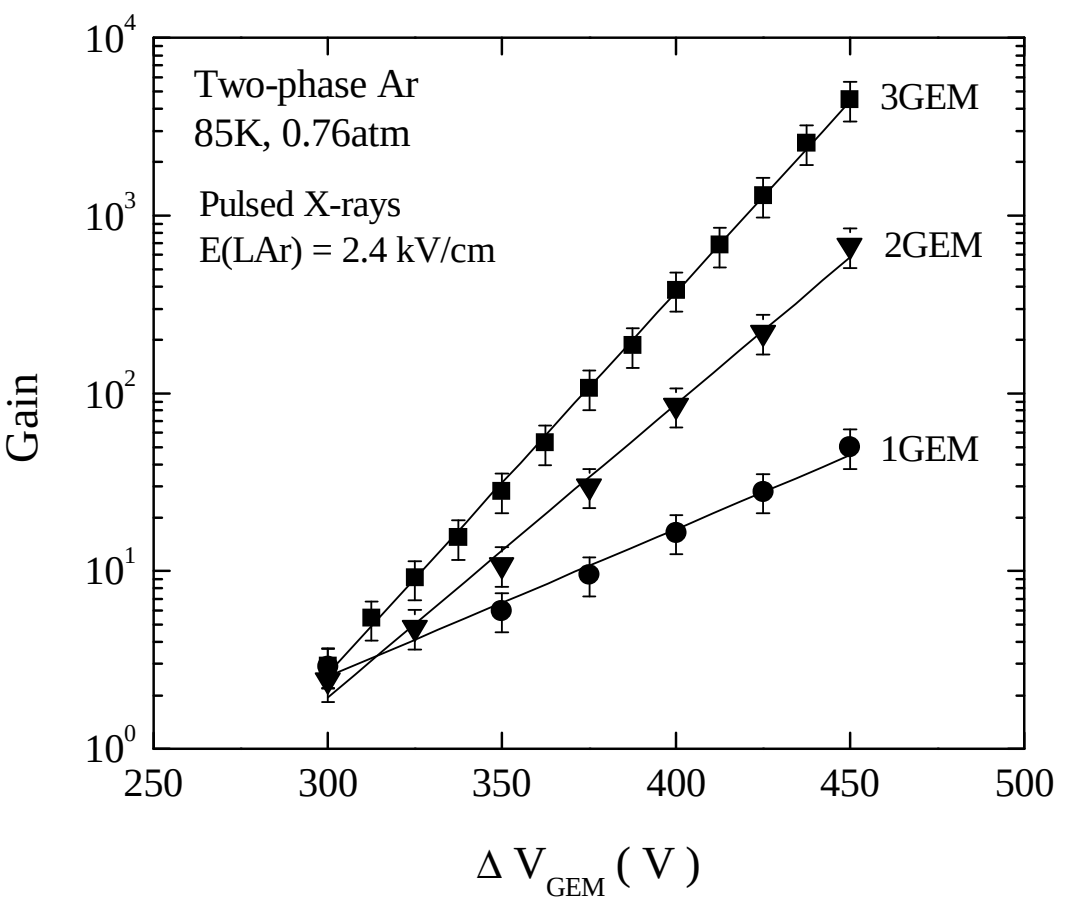

Fig.5 Gain-voltage characteristics of the single-, double- and triple-GEMs measured using pulsed X-rays, in two-phase Ar at 85 K and 0.76 atm. The maximum gains for single- and double-GEMs were not reached.

Figs. 6 and 7 illustrate a detector response to 60 keV $\gamma$-rays from a $^{241}$Am source. Fig. 6 shows pulse-height spectra from the triple-GEM at a gain of 2500 at two different electric fields in the liquid. A distinct peak corresponding to the absorption of 60 keV photons and a tale, presumably corresponding to the events with scattered photons, are seen in the figure. The effect of the extraction field is well pronounced: peak positions are in good correspondence with electron emission probability obtained from Fig. 3. The energy resolution obtained by the Gauss fit of the peak is 37% (FWHM). It is defined mostly by pressure variations during the measurements (which were lasting typically 10-20 minutes). We expect that the energy resolution will be improved in the detectors having better temperature/pressure stabilization.

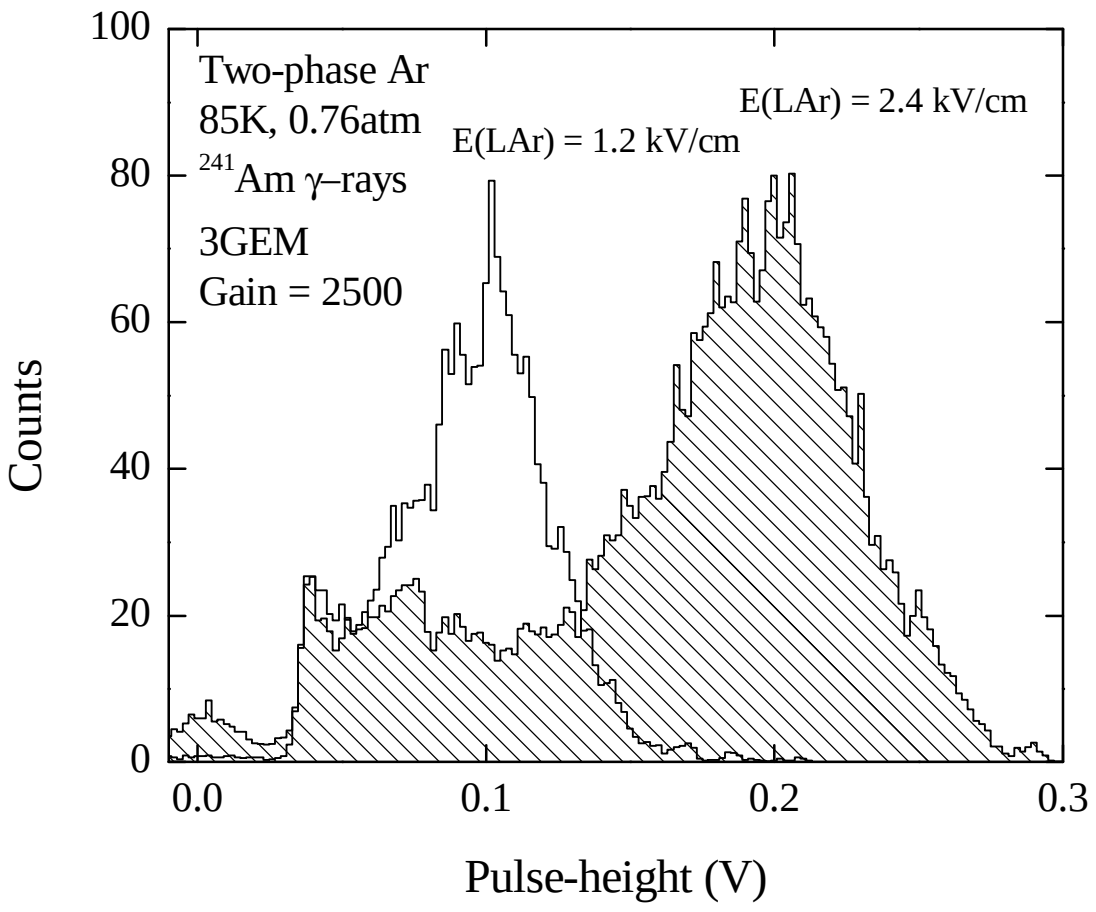

Fig.6 Pulse-height spectra from the triple-GEM induced by 60 keV $\gamma$-rays from a $^{241}$Am source, at a gain of 2500 at two different electric fields in the liquid, in two-phase Ar at 85 K and 0.76 atm.

The anode signals from the triple-GEM induced by 60 keV $\gamma$-rays, in two-phase Ar, are shown in Fig. 7. At moderate gain (700) the pulse shape is almost triangular, while at higher gain (2500) the primary signal is accompanied by a secondary signal. The secondary signals at comparable gains and time scales were observed earlier in gaseous Ar [11]: they were explained by photon feedback between the last GEM and the copper cathode, enhanced at cryogenic temperatures. In the present version of the setup, the cathode could hardly provide the feedback since it had a small effective area. Most probably, the secondary signals were caused by photon feedback between the GEM elements themselves.

A detector response to pulsed X-rays is illustrated in Fig. 8: anode signals and a pulse-height spectrum from the triple-GEM are shown. The amplitude resolution is 14.0% (FWHM). Here the measurement time was short, of a few seconds, resulting in that the contribution of pressure variations to the resolution was negligible. In that case the amplitude resolution is determined by X-ray photon statistics.

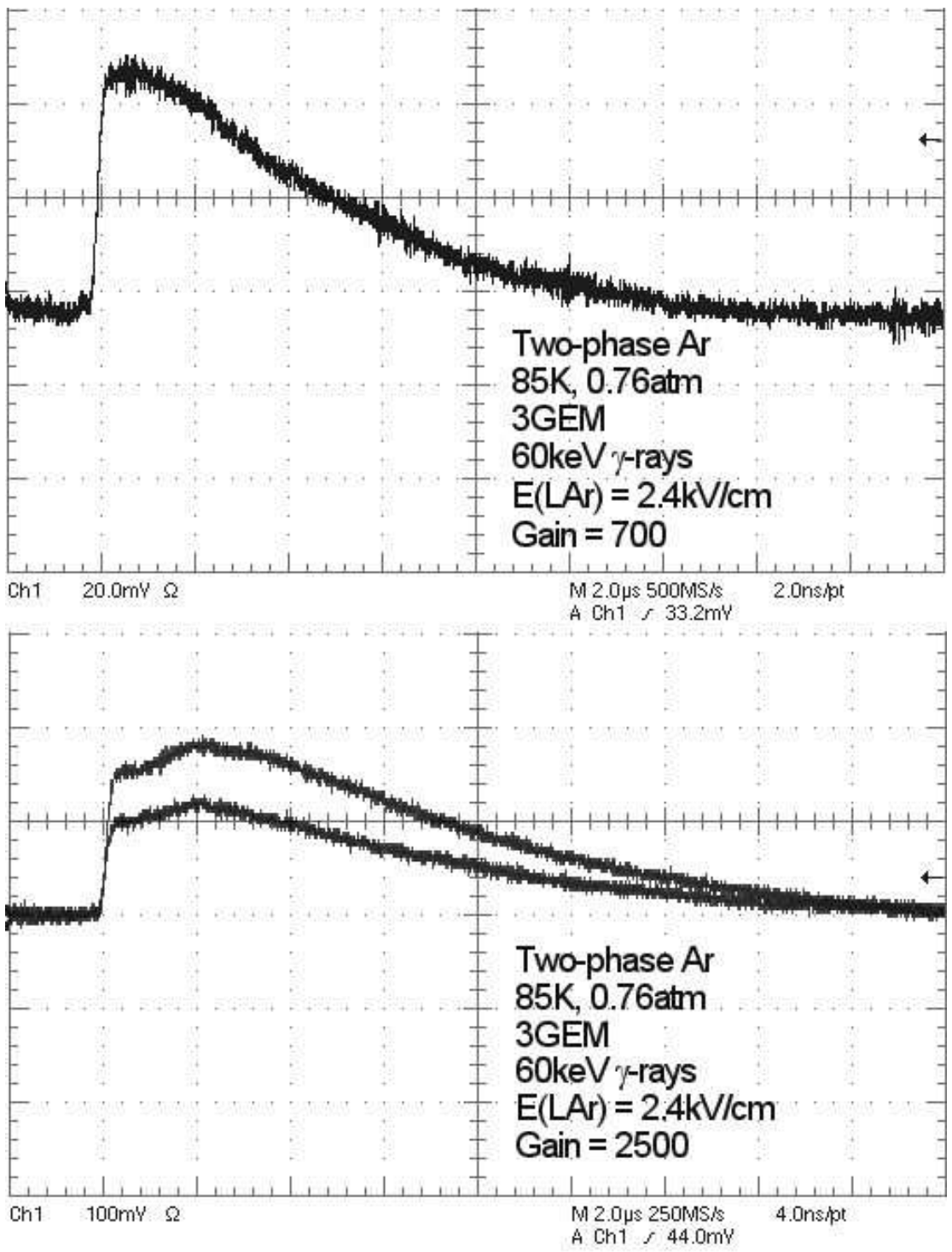

Fig.7 Typical anode signals from the triple-GEM induced by 60 keV γ-rays from a $^{241}$Am source, at a gain of 700 (top) and 2500 (bottom), in two-phase Ar at 85 K and 0.76 atm.

Indeed, accounting for the pulse height of the calibration signal, amplifier sensitivity and fraction of primary electrons escaping recombination (which is taken to be 50% for 40 keV photons [21]), the energy deposited in the liquid by an X-ray pulse is estimated to be 16 MeV, which roughly corresponds to 400 X-ray photons absorbed. Thus, the contribution of photon statistics to the resolution is $2.35/\sqrt{400}=12\%$ which is close to that observed in experiment. This fact indicates that the Ar purity was high enough to collect the charge from the whole of the liquid layer, i.e. that the electron drift path in liquid Ar before attachment was larger than 3 mm.

A detector response to β-particles from a $^{90}$Sr source is illustrated in Fig. 9: pulse-height spectra from the double-GEM at a gain of 700 are shown. The counting rate was of the order of 10 kHz. The range of a 1 MeV electron in liquid Ar is 3 mm. That means that half of the β-particles would not stop in the liquid, resulting in that the β-spectrum measured in the detector would be distorted. Nevertheless, one can see that the shape of the β-spectrum is reflected in the pulse-height spectra in general and that the dependence on the extraction field is well pronounced, similarly to that observed in Fig. 6.

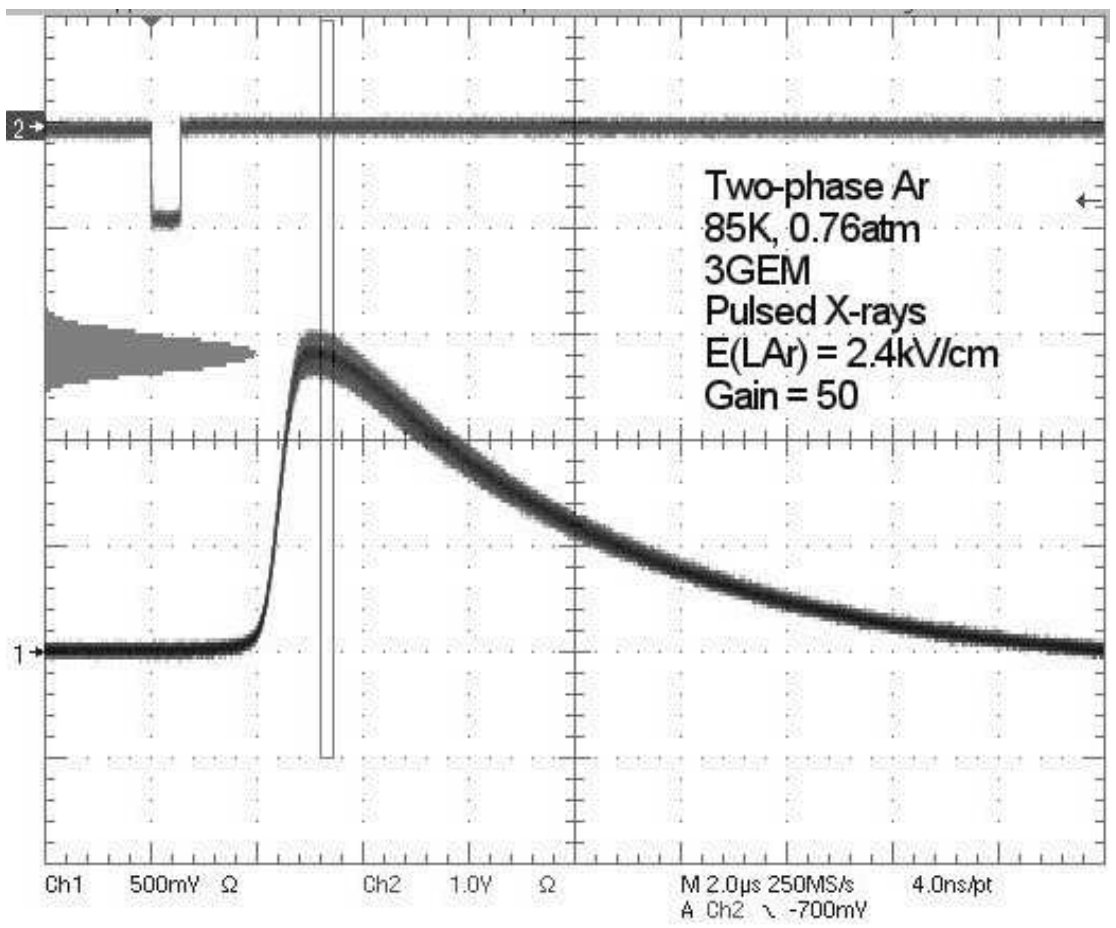

Fig.8 Anode signals from the triple-GEM induced by pulsed X-rays in two-phase Ar at 85 K and 0.76 atm, at a gain of 50. A pulse-height spectrum, on the left, and trigger signal from an X-ray tube, on the top, are also shown. The energy deposited in the liquid by an X-ray pulse is estimated to be about 16 MeV, corresponding to about 400 X-ray photons absorbed.

In addition, Fig. 9 gives an idea of the response of two-phase Ar avalanche detectors to minimum ionizing particles. For example, it follows from Fig. 9 that the average amplitude of the signals induced by β-particles, reduced to the gain value of Fig. 6, would be by an order of magnitude larger than for those induced by 60 keV γ-rays. This fact indicates that the detector response to minimum ionizing particles may produce a serious background problem in low-background experiments.

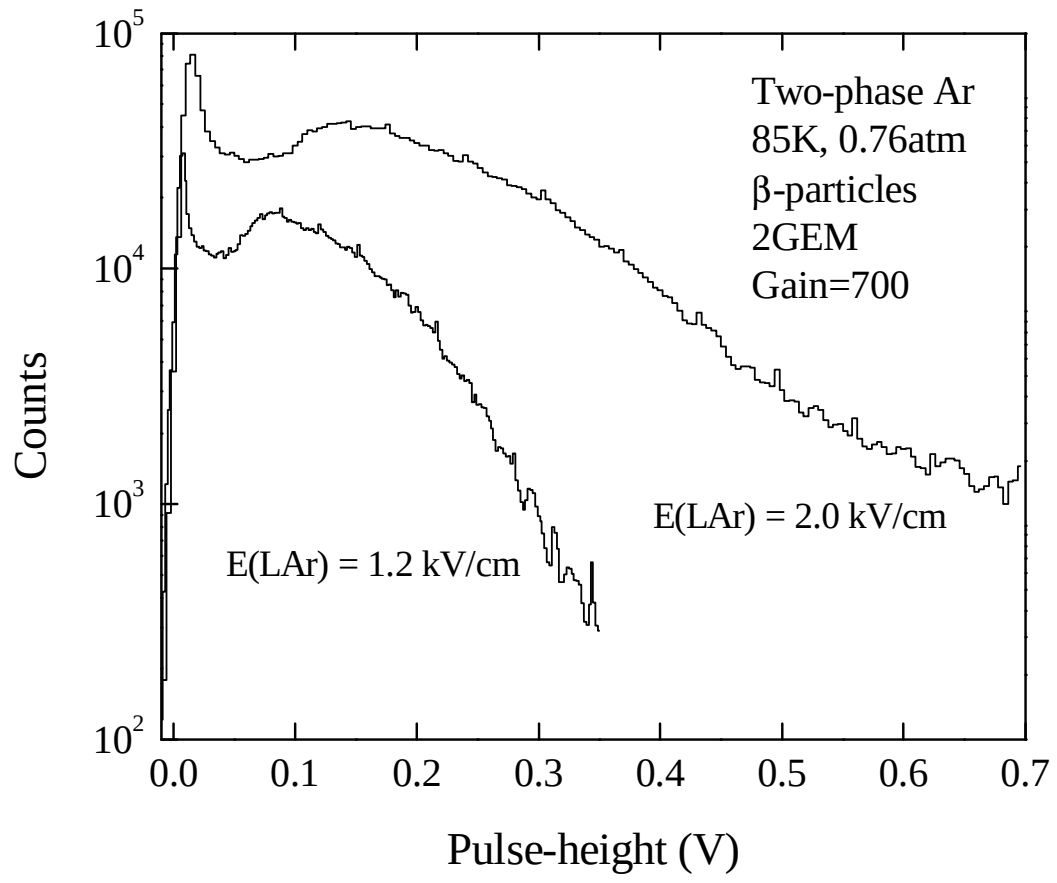

Fig.9 Pulse-height spectra from the double-GEM induced by β-particles from a $^{90}Sr$ source, at a gain of 700 at two different electric fields in the liquid, in two-phase Ar at 85 K and 0.76 atm.

The operation of the two-phase Ar avalanche detector, using the triple-GEM multiplier, was rather stable and reproducible, in particular at the point of the best pressure stabilization. No gain instabilities were observed for at least 6 hours; during this period the detector was under voltage and the height of the pulse from the triple-GEM, induced by pulsed X-rays or β-particles, was measured at regular intervals.

## 4. Two-phase Xe avalanche detector

Fig. 10 shows an evolution of the cathode gap capacitance as a function of the Xe gas/vapor pressure during the cooling and heating procedures. During the cooling procedure in the pressure range of 0.9-1.8 atm, the detector operates in a two-phase mode: the presence of the liquid Xe condensate at the chamber bottom is confirmed by an increase of the gap capacitance and its fluctuations, similarly to that observed in the two-phase Ar detector (Fig. 2).

When cooling further (at a pressure below 0.9 atm) the capacitance fluctuations decrease stepwise indicating a transition from the liquid to solid state, in contrast to the Ar detector. At the end of the cooling cycle, the capacitance gradually returns to the initial value, indicating a disappearance of the condensate at the chamber bottom, similarly to that observed in the two-phase Kr detector [12]. This takes place due to a sublimation of solid Xe at the chamber bottom which finally condenses at the coldest part of the chamber, i.e. at the heat exchanger. At the beginning of the heating cycle, in the pressure range of 0.2-0.8 atm, there is no condensate at the chamber bottom at all.

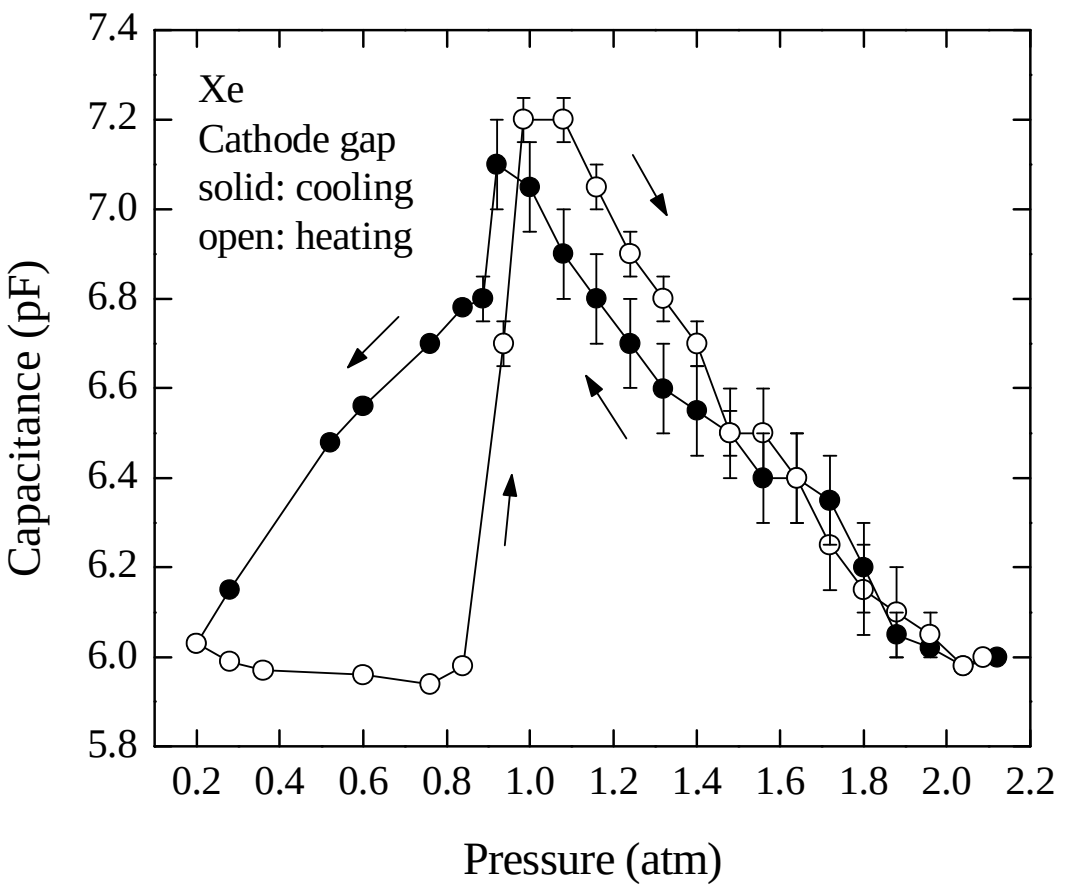

Fig.10 Cathode gap capacitance as a function of the gas/vapor pressure during cooling and heating procedures in Xe.

The region of the cooling/heating cycle between 0.9 and 1.8 atm was used to study the detector performance in a two-phase (liquid-gas) mode. The region of the heating cycle between 0.2 and 0.8 atm was used to study the GEM performance in a “single-phase” (gaseous) mode, at cryogenic temperatures.

The two-phase Xe detector based on GEM multipliers was successfully operated in an avalanche mode at moderate gains. Fig. 11 illustrates an electron emission characteristic in the two-phase Xe detector measured using pulsed X-rays. In general, the emission characteristic is in agreement with the data reported in the literature [20]: the emission starts at about 2 kV/cm and its probability keeps growing at fields higher than 4 kV/cm.

The liquid and absorbed in the vapor phase. In contrast to Ar, this component is observed in two-phase Xe because the electron drift path in liquid Xe was much smaller than that in Ar, resulting in a reduction of the electron emission component. The electron drift path in liquid Xe is estimated to be 0.6 mm, taking into account the relative contributions of the components in Fig. 11 and X-ray absorption coefficients in liquid and gaseous Xe.

Fig.12 shows gain-voltage characteristics of the triple-GEM in two-phase Xe at a vapor pressure of 0.88 and 0.99 atm, measured using pulsed X-rays. In Fig. 4 the gain characteristic is compared to those in two-phase Ar and Kr. One can see that the maximum gain in two-phase Xe is about 200, which is somewhat lower than that in Kr and substantially lower than that in Ar.

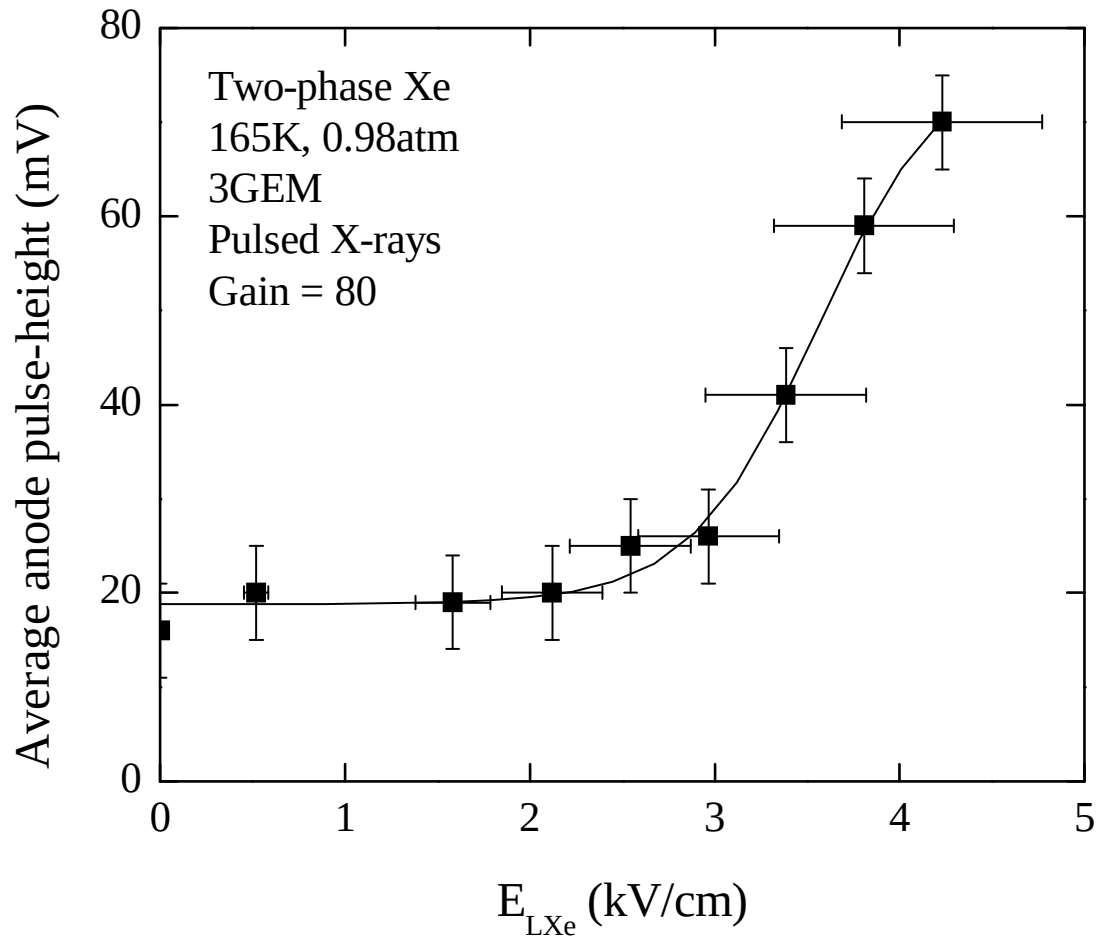

Fig.11 Electron emission characteristic in the two-phase Xe avalanche detector at 165 K and 0.98 atm. Average anode pulse-height from the triple-GEM induced by pulsed X-rays is shown as a function of the electric field in liquid Xe. The triple-GEM gain is 80.

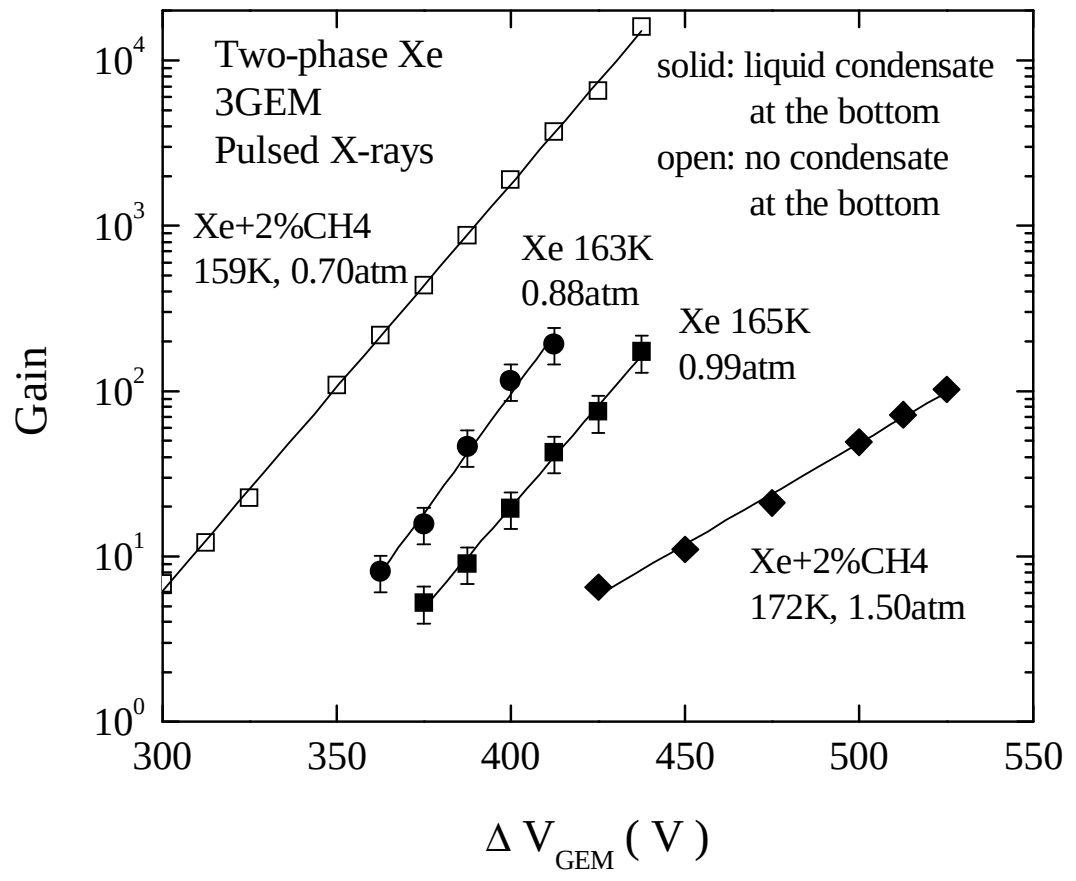

Fig.12 Gain-voltage characteristics of the triple-GEM, measured using pulsed X-rays, in two-phase Xe and two-phase Xe+$CH_4$, when there is a liquid condensate at the chamber bottom, and in gaseous Xe+$CH_4$, when there is no condensate at the bottom. The appropriate temperatures, pressures and $CH_4$ concentrations are indicated. In the two-phase mode, the electric field in liquid Xe is 4.0 kV/cm. The maximum gains are limited by discharges.

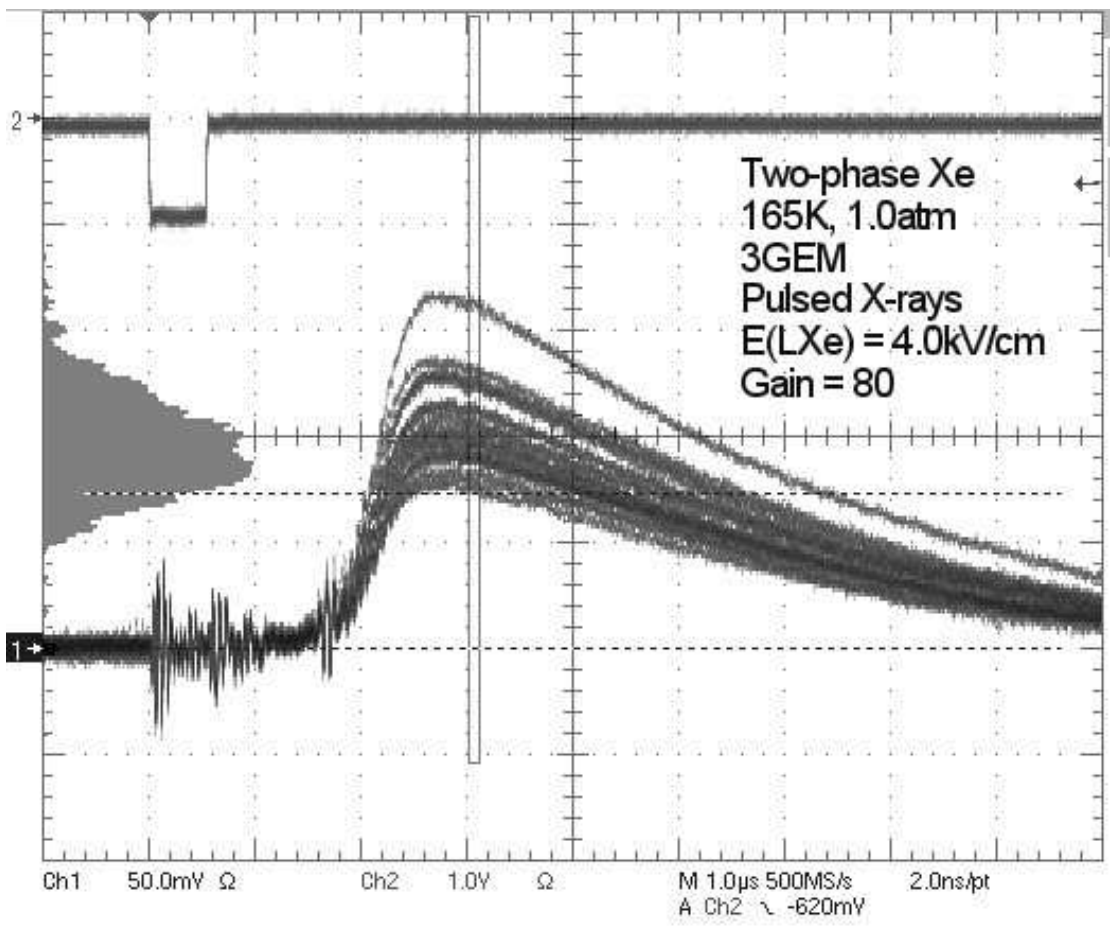

Fig.13 Anode signals from the triple-GEM induced by pulsed X-rays in two-phase Xe at 165 K and 1.0 atm, at a gain of 80. A pulse-height spectrum, on the left, and trigger signal from an X-ray tube, on the top, are also shown.

Along with the component depending on the electric field and induced by electron emission from the liquid, there is a component which is independent of the electric field. It is presumably induced by X-rays passed through

A detector response to pulsed X-rays is illustrated in Fig. 13: anode signals and a pulse-height spectrum from the triple-GEM are shown. The amplitude of the signal is lower (by a factor of 30 if normalized to the gain) and the amplitude resolution is worse (65%, FWHM) compared to the two-phase Ar detector (Fig. 8), obviously due to the effect of electron attachment in liquid Xe. The “detectable” energy which might be attributed to this resolution, i.e. the energy corresponding to the amount of ionization reached the triple-GEM per X-ray pulse, is estimated to be 200 keV.

The operation stability of gas multipliers in two-phase Xe is of particular importance. Indeed, it was observed that the period of stable operation of the Micromegas and proportional chamber in saturated Xe vapor did not exceed half an hour [19,22]. This instability was believed to be due to Xe vapor condensation within the gas multiplier. In the present work, the operation stability of the triple-GEM in two-phase Xe was measured for a period of half an hour at a gain of 80 under irradiation with β-particles. During this period the triple-GEM operation was relatively stable and the gain variations observed (of a factor of 2) were correlated to the vapor pressure variations (of about 5%). It should be remarked that this result is rather preliminary, due to the large pressure variations induced by insufficient temperature stabilization.

We studied the possibility to increase the maximum gain in the two-phase Xe detector by adding a quenching gas, namely methane. In that case the mixture Xe+2.1%$CH_4$ was prepared in the bottle at room temperature. Since the maximum dissolution level of $CH_4$ in liquid Xe was found to be 1.9% [23], the $CH_4$ concentration in Xe vapor in the two-phase mode might be taken as approximately 2%. Fig. 12 shows gain-voltage characteristics of the triple-GEM in Xe+$CH_4$ mixture in the two-phase mode, when there is a liquid condensate at the chamber bottom, and in the gaseous mode, when there is no condensate at the bottom. At cryogenic temperatures, in the gaseous mode the gain can easily exceed $10^4$, while in the two-phase mode the maximum gain does not exceed 200, i.e. is practically the same as in two-phase Xe. This result indicates that just the operation in the two-phase mode imposes a principal limit on the maximum gain.

## 5. Conclusions

The performance of two-phase (liquid-gas) Ar and Xe avalanche detectors, based on GEM structures and operated in an electron-avalanching mode, have been studied for the first time. Emission, gain, energy resolution and stability characteristics of the detectors were studied.

The characteristics obtained in the two-phase Ar avalanche detector are rather promising. Rather high gains, reaching 5000, were obtained in the detector using a triple-GEM multiplier operated in saturated Ar vapor above the liquid. In Ar, the probability of electron emission through the liquid-gas interface reaches a plateau at a relatively low electric field in the liquid, as compared to Kr and Xe, namely at 2.5 kV/cm. The signals induced by X-rays, β-particles and 60 keV γ-rays were successfully recorded. In the latter case, the energy resolution obtained was 37% (FWHM); it is defined by pressure variations during the measurements and should be improved in the detector with better temperature stabilization. A stable operation of the triple-GEM multiplier in saturated vapor above the liquid was observed for at least 6 hours. There were no charging-up effects observed, even at anode current densities as large as 2.5 $nA/mm^2$.

The GEM-based two-phase Xe avalanche detector was successfully operated at moderate gains. The electric fields needed for efficient electron emission from liquid Xe are by a factor of 2-3 higher, compared to Ar. The maximum gain of the triple-GEM in saturated Xe vapor above the liquid is about 200. Adding a few percents of methane to saturated Xe vapor did not increase the maximum gain. A relatively stable operation of the detector was observed for half an hour when irradiated with β-particles. However, further measurements are needed in order to understand the instability mechanism in two-phase Xe avalanche detectors observed by other authors. Low gain values obtained in such detectors are presumably too small for dark matter search experiments. Accordingly, the ways to increase the gain should be found.

The results obtained, in particular in Ar, are relevant for developing two-phase avalanche detectors for dark matter searches, coherent neutrino scattering, PET and digital radiography. We expect to improve the temperature stabilization, gain, energy resolution and liquid purity characteristics in new generations of two-phase avalanche detectors. They are being developed in our laboratory.

## Acknowledgements

The research described in this publication was made possible in part by Award RP1-2550-NO-03 of the U.S. Civilian Research & Development Foundation for the Independent States of the Former Soviet Union (CRDF) and in part by Award 04-78-6744 of INTAS Grant. This work has been partially motivated by the possible application in cryogenic two-phase detectors for solar neutrino and dark matter detection. We are indebted to Prof. W. Willis and Drs. J. Dodd and M. Leltchouk, of the Columbia University, and Prof. N. Spooner of the Sheffield University for having suggested these applications.